\documentclass[aps,prl,reprint,twocolumn,floatfix]{revtex4-2}
\usepackage[utf8]{inputenc}
\usepackage[english]{babel}
\usepackage[T1]{fontenc}
\usepackage{amsmath}
\usepackage{hyperref}
\usepackage{graphics}
\usepackage{lipsum}
\usepackage{units}
\usepackage{upgreek}
\usepackage{units}

\newcommand{\ee}{\mathrm{e}}
\newcommand{\ii}{\mathrm{i}}
\newcommand{\alphat}{\alpha^{(\mathrm{t})}(\omega,\mathbf{u}_1,\mathbf{u}_2)}
\newcommand{\alphar}{\alpha^{(\mathrm{r})}(\omega,\mathbf{u}_1,\mathbf{u}_2)}
\newcommand{\aIP}{\alpha_{\mathrm{IP}}}
\newcommand{\aOP}{\alpha_{\mathrm{OP}}}
\newcommand{\taIP}{\tilde\alpha_{\mathrm{IP}}}
\newcommand{\taOP}{\tilde\alpha_{\mathrm{OP}}}

\newcommand{\He}{$^4$He$^+$}
\newcommand{\Be}{$^9$Be$^+$}
\newcommand{\Al}{$^{27}$Al$^{+}$}
\newcommand{\Ca}{$^{40}$Ca$^+$}
\newcommand{\Shalb}{$^2$S$_{1/2}$}
\newcommand{\Snull}{$^{1}$S$_{0}$}
\begin{document}

\title{A scheme for quantum-logic based transfer of accuracy in polarizability measurement for trapped ions using a moving optical lattice}
\date{\today}
\author{Fabian Wolf}
\affiliation{Physikalisch-Technische Bundesantalt, 38116 Braunschweig, Germany}
\email{fabian.wolf@ptb.de}

\begin{abstract}
  Optical atomic clocks based on trapped ions suffer from systematic frequency shifts of the clock transition due to interaction with blackbody radiation from the environment. 
  These shifts can be compensated if the blackbody radiation spectrum and the differential dynamic polarizability is known to a sufficient precision. 
  Here, we present a new measurement scheme, based on quantum-logic that allows a direct transfer of precision for polarizability measurements from one species to the other.
  This measurement circumvents the necessity of calibrating laser power below the percent level, which is the limitation for state-of-the-art polarizability measurements in trapped ions.
  Furthermore, the presented technique allows to reference the polarizability transfer to hydrogen-like ions for which the polarizability can be calculated with high precision.
  \end{abstract}

\maketitle
  Advances in the precision of atomic clocks~\cite{ludlow_optical_2015} have enabled a new class of fundamental physics tests such as probing for a possible variation of fundamental constants~\cite{godun_frequency_2014,huntemann_improved_2014} certain types of dark matter candidates~\cite{wcislo_new_2018} and violations of local Lorentz symmetry~\cite{sanner_optical_2019}. 
  One class of systems that have proven successful in the past are optical clocks based on transitions in single trapped ions.
  In these systems, uncertainties in the $10^{-18}$~\cite{huntemann_single-ion_2016} range and below~\cite{brewer_27+_2019} have been demonstrated.
  A significant contribution to these uncertainties stems from blackbody radiation induced light shifts of the clock transition.
  This effect can be compensated if the temperature environment of the ion~\cite{Dolezal_analysis_2015} and the differential dynamic polarizability are known to a sufficient precision.

	The differential polarizability of the states involved in the clock transition is usually inferred by measuring the shift of the clock transition under illumination with infrared laser radiation.
	This scheme has been implemented for the ytterbium~\cite{baynham_measurement_2018,huntemann_single-ion_2016}, aluminum~\cite{brewer_27+_2019}, and lutetium~\cite{arnold_blackbody_2018,arnold_dynamic_2019} single ion clock transitions.
	It turned out, that the determination of the laser intensity poses a major challenge for achieving sub-percent uncertainties using techniques based on laser illumination.
	An alternative approach applicable to transitions with a negative differential static scalar polarizability has been first demonstrated for the $^{88}$Sr$^{+}$-clock~\cite{dube_high-accuracy_2014} later applied to $^{40}$Ca$^+$~\cite{huang_40mathrmca_2019} and recently shown to be feasible with $^{138}$Ba$^+$~\cite{barrett_polarizability_2019}.
	The negative static scalar polarizability allows to cancel the trap drive-induced ac Stark shift with the micromotion-induced time-dilation shift by tuning the trap-drive frequency to a so-called \textit{magic} value.
	By measuring this \textit{magic} trap-drive frequency, the differential static scalar polarizability can be inferred.
	This scheme has proven to achieve precisions on the sub-percent level.
	
	Recently, it was proposed~\cite{barrett_polarizability_2019} to combine two clock species of which one offers a transition with a well-known differential polarizability.
	This ion can serve as a power reference for the calibration of the laser power to measure the second ion's polarizability.
	Here, we propose an alternative scheme in which a quantum logic operation is implemented to transfer the accuracy of polarizabilites between two states of different ions. 
	This allows an efficient cancellation of effects from drifts in the laser intensity.
	Furthermore, the measurement of absolute polarizabilities rather than differential polarizabilities allows to use the groundstate of hydrogen-like systems such as $^4$He$^{+}$ as a reference ion, where the polarizability can be computed to a very high level of precision~\cite{szmytkowski_static_2016}. While the method outlined in reference \cite{barrett_polarizability_2019} offers a direct determination the differential polarizability, it involves the necessity of probing the clock transition. In contrast, the approach presented here has the potential to extend its capabilities to measure differential polarizabilities, as discussed in the outlook, without the need to measure the clock transition. In this technique, the clock laser serves solely to prepare the ion in one of the two clock states.   

\begin{figure}[htpb]
	  \centering
	  \includegraphics{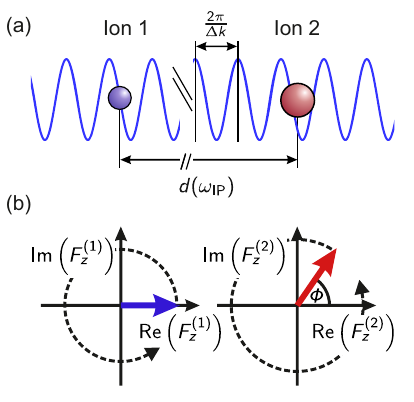}
	  \caption{(a) The two ions are separated along the $z$-axis by distance $d$. Because of the spatial dependence of the intensity with characteristic length scale $2\pi/\Delta k$, the ions sample the oscillating force with different phases $\phi$. The amplitude of the oscillating force depends on the dynamic polarizability. Subfigure (b) shows pointer diagrams of the forces on the individual ions.}
	  \label{fig:lattice}
	\end{figure}

	In contrast to previously demonstrated schemes where a single infrared laser beam was used to imprint the ac-Stark shift on the clock ions, we propose to irradiate the ions with two infrared laser beams from the same laser source.
 The interference of these beams will lead to a spatial modulation of the light intensity 
	which allows to employ the light shift to couple to the motional state of the two-ion crystal.	
	Since the ions sample different positions of the spatially varying light field, the optical forces will add up with a phase, given by their relative positions~(see figure~\ref{fig:lattice}) and their individual coupling to the excited mode~\cite{wolf_non-destructive_2016}.
	Exciting different modes of the two ion crystal allows to probe different independent linear combinations of the two forces acting on the ion.
	Comparing these excitation allows to infer the amplitudes of the individual forces and therefore the ratio of polarizabilities of the ions' occupied states.
	Given that the light field can be made sufficiently spatially homogeneous across the distance of a few micrometers between the ions and that most properties of the light field are constant on a time scale of an individual experiment, interleaved probing of the different modes renders the scheme robust against effects related to comparably slow intensity and mode profile fluctuations of the light field.
      The proposed experiment, in its simplest implementation, is based on a mixed species two-ion crystal confined in a linear Paul trap.
      One ion serves as the reference ion~(RI) and the other as the target ion~(TI). 
      The goal of the measurement is to infer the polarizability of the target ion by comparing it to the reference ion.
      Therefore, the polarizability of at least one state in the reference ion should be known to the level of precision that is desired for the polarizability determination of the target ion.
      This state should be long lived with respect to the experimental cycle and techniques are required to populate it with high fidelity.
      The target ion is prepared in the state whose polarization is to be determined.
      At least one of the ions needs a suitable level structure for ground state cooling, a qubit for storage of information and qubit-readout capabilities.
      It can be either the reference ion or the target ion. Alternatively, also quantum logic schemes~\cite{schmidt_spectroscopy_2005} where these tasks are distributed are possible.
      In addition to the cooling and qubit-manipulation laser systems an additional infrared laser is required, that will imprint the ac-Stark shift.
      The infrared laser light is split into two beams that can be shifted in frequency with respect to each other by a few megahertz and are sent along the trap axis in a counter-propagating configuration, forming a moving optical lattice~\cite{ding_microwave_2014,wolf_non-destructive_2016,najafian_identification_2020} 

      If the relative detuning of the two laser beams is resonant with one motional mode frequency of the linear ion chain, the resulting dynamics is described by the unitary operator
      \begin{equation}
	D(\alpha_k)=\ee^{\alpha_k \hat a_k^\dagger - \alpha_k^* \hat a_k},
	\label{eq:Disp}
      \end{equation}
      with the creation and annihilation operator $\hat a_k^\dagger$ and $\hat a_k$ for motional mode $k$.
      This operator is known as the displacement operator that creates a coherent state in motional mode $k$ with amplitude
      \begin{equation}
	\alpha_k=\frac{\ii}{\hbar} \sum_j \eta_k^{(j)} \varepsilon_1\varepsilon_2^* \alpha^{(j)}(\omega,\mathbf{u}_1,\mathbf{u}_2)\ee^{-\ii\phi_j}t_\mathrm{F}, 
	\label{eq:Displacement}
      \end{equation}
      where we introduced the Lamb-Dicke parameter $\eta^{(j)}_k=\Delta k_z \beta^{(j)}_k \sqrt{\frac{\hbar}{2m_j\omega_k}}$with the mass of the $j$-th ion $m_j$, mode frequency $\omega_k$ and $\beta^{(j)}_k$ the transformation matrix element for the $j$-th ion between the lab frame and the shared motional modes. $t_k$ denotes the interrogation time of the moving optical lattice. 
      In general the polarizability $\alpha^{(j)}(\omega,\mathbf{u}_1,\mathbf{u}_2)$ depends on the laser wavelength $\omega$ and the polarization vectors $\mathbf{u}_1$ and $\mathbf{u}_2$.
      For a two-ion crystal we can rewrite this system of equations to determine the polarizability of the target ion with respect to the reference ion~\cite{supp_mat} and find
      \begin{equation}
	|\alphat|=|\alphar| \Gamma\left(\mu,r ,\Phi\right),
	\label{^k}
      \end{equation}
      where $\Gamma$ is a function, that only depends on the mass ratio $\mu=m_\mathrm{t}/m_\mathrm{r}$ between the target ion and the reference ion, the ratio of displacements created on the in-phase (IP) and out-of-phase (OP) motion $r=|\aIP|/|\aOP|$ and the phase difference $\Phi=\phi_\mathrm{r}-\phi_\mathrm{t}$ of the oscillating force sampled by the two ions. 
      Assuming that we can control the phase close to $\Phi=0$ by monitoring the trapping frequency and angular alignment of the trap axis and laser directions, we can write
      \begin{equation}
	\Gamma\left(\mu,r ,\Phi\right)=\Gamma\left(\mu,r \right)\left( 1+\mathcal{O}(\Phi^2) \right),
	\label{<+label+>}
      \end{equation}
      with 
\begin{equation}
	\Gamma\left( \mu,r\right)=
	\sqrt{\mu}\frac{\pm\beta_{\mathrm{IP}}^{(r)}-\beta_{\mathrm{OP}}^{(r)}\sqrt{\frac{\omega_{\mathrm{IP}}}{\omega_{\mathrm{OP}}}}r}{\beta_{\mathrm{OP}}^{(t)}\sqrt{\frac{\omega_{\mathrm{IP}}}{\omega_{\mathrm{OP}}}}r\mp\beta^{(t)}_{\mathrm{IP}}}.
	\label{eq:main}
      \end{equation}
      If the upper or lower sign has to be used depends on the polarizabilities and mass ratio of the target and reference ion. Details on how to choose the sign are given in the supplementary information.
      The displacement amplitudes $|\aIP|$ and $|\aOP|$ can be determined experimentally~\cite{wolf_motional_2019} and all other parameters except for the polarizabilities only depend on the mass ratio of the two ions~\cite{wubbena_sympathetic_2012}, which can be determined very precisely in Penning trap experiments~\cite{blaum_high-accuracy_2006}.

\begin{figure}[htpb]
	\centering
	\includegraphics{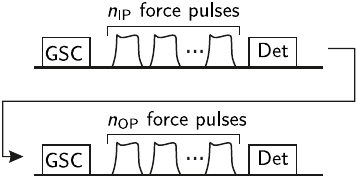}
	\caption{Sketch of the experimental sequence. The ions are prepared in the initial motional state e.g. by ground state cooling~(GSC) in both modes. Afterwards a sequence of $n_{IP/OP}$-pulses with the moving optical lattice is applied to coherently excite the motion which is then detected~(Det) by mapping the motional state onto the read-out ion's internal state. The sequence is repeated for the other motional mode. 
	}
	\label{fig:sequence}
      \end{figure}

      The proposed experimental sequence for the determination of the displacement ratio $|\aIP|/|\aOP|$ starts with preparing the two ion crystal in the initial motional state.
      In the simplest implementation this is done by ground-state cooling of both axial motional modes. 
      Afterwards, the infrared moving optical lattice is applied to displace the motional wave package of one mode in phase space.
      In the end the overlap of the final state with the initial state is measured to infer the induced displacement amplitude.
      For an initially ground state cooled ion crystal, this can by done by STIRAP or RAP on the red-sideband transition~\cite{gebert_detection_2016,gebert_corrigendum_2018}.
      A quantum-enhanced version of the experiment could employ squeezed states~\cite{burd_quantum_2019}, Fock states~\cite{wolf_motional_2019} or Schr\"odinger cat states~\cite{hempel_entanglement-enhanced_2013} for faster averaging of quantum projection noise.
      Both axial modes should be interrogated in an interleaved fashion, such that influences from drifts in laser intensity cancel out.
      Furthermore, for the described displacement measurement it is advantageous to measure displacements on the order of one to operate close to the working point of maximum sensitivity~\cite{wolf_motional_2019}. 
      Therefore, it is convenient to apply the displacement in a pulsed fashion (see figure~\ref{fig:sequence}) with variable pulse numbers $n_\mathrm{IP}$ and $n_{\mathrm{OP}}$ but fixed pulse length $t_\mathrm{pulse}$ such that the measured displacements fulfill $n_{\mathrm{IP}}|\aIP|\approx 1$ and $n_{\mathrm{OP}}|\aOP|\approx 1$. 
    An additional advantage of this scheme is that errors due to pulse shape distortions are suppressed as long as the pulse shapes are identical for the interrogation of both modes.

\begin{figure}[htpb]
	\centering
	\includegraphics{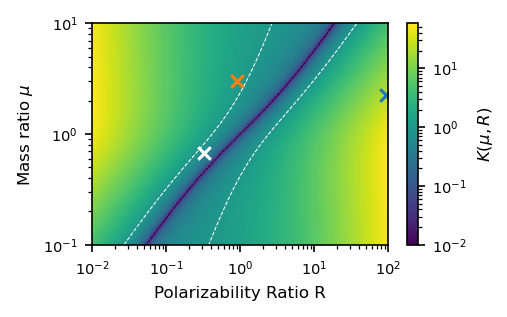}
	\caption{Mass ratio $\mu$ and polarizability ratio $R$ dependent sensitivity factor. The graph shows the sensitivity factor $K(\mu)$ that relates the relative measurement uncertainty in the displacement ratio $\Delta r/r$ to the relative uncertainty in polarizability determination $\Delta \alphat/\alphat$. The region between/outside the dashed lines corresponds to a sensitivity factor below/above 1. The crosses correspond to physical systems. Blue: \He~\Shalb (reference) - \Be~\Shalb (target); Orange: \Be~\Shalb (reference) - \Al~\Snull (target); White: \Ca~\Shalb (reference) - \Al~\Snull (target).}
	\label{fig:sensitivity}
      \end{figure}

      In order to get an estimate of the achievable sensitivity, we can define a mass ratio-dependent uncertainty propagation factor which relates the fractional uncertainty in the displacement amplitude determination to the relative uncertainty in the polarizability ratio by
      \begin{equation}
	\frac{\Delta R}{R}=K(\mu, R)\frac{\Delta r}{r},
	\label{eq:sensitivityK}
      \end{equation}
      where $R = |\alphat|/|\alphar|$ is the ratio of polarizabilities of the target ion and the reference ion.
      In the supplementary information it is shown that the uncertainty propagation factor is given by
      \begin{align}
	K(\mu, R) = \frac{\mu\left(1+2R\right) - R \left( 2+R \right)}{2R \sqrt{\mu^2-\mu+1}}.
	\label{<+label+>}
      \end{align}
      Figure~\ref{fig:sensitivity} shows the uncertainty propagation factor $K(\mu)$ in dependence of the mass ratio $\mu$ and the polarizability ratio $R$. 
      It can be seen that $K(\mu,R)$ vanishes for $R=\mu-1+\sqrt{\mu^2-\mu+1}$, which is the regime, where the forces on the two ions cancel exactly for $\Phi=0$ and the out-of-phase motion cannot be excited. 
      The region where $K(\mu,R)$ is smaller than one, i.e. the relative uncertainty in the polarization determination is smaller than the relative uncertainty of the displacement amplitude ratio measurement, lies between the dashed lines. 
Eventually the achievable accuracy is limited by the knowledge of the polarizability of the reference ion and the uncertainty in the measured polarizability of the target ion and is given by
  \begin{equation}
    \frac{\Delta \alphat}{\alphat} = \sqrt{K(\mu,R)\frac{\Delta r}{r}+\frac{\Delta \alphar}{\alphar}}.
    \label{<+label+>}
\end{equation}
     
\begin{table}[htpb]
	\caption{\label{tab:pol} Literature values for dipole polarizabilities. 
All values were determined theoretically. For helium and aluminum static polarizabilities are given and extracted from reference~\cite{szmytkowski_static_2016} and \cite{yu_finite-field_2013}, respectively. For beryllium the value for $0.04$ (a.u.) (corresponds to around $\unit[1.1]{\upmu m}$ laser wavelength) is given from reference~\cite{tang_dynamic_2010}.}
	\begin{ruledtabular}
	  \begin{tabular}{@{}rll@{}}
	    Species & State &Dipole polarizability (A$^2$s$^4$kg$^{-1}$)\\
	    $^{4}$He$^{+}$&$^{2}\mathrm{S}_{1/2}$&$4.636161773523698(462)\times10^{-42}$~\cite{szmytkowski_static_2016}\\
	    $^{9}$Be$^{+}$&$^{2}$S$_{1/2}$&$4.3559(66)\times10^{-40}$~\cite{tang_dynamic_2010}\\
	    $^{27}$Al$^{+}$ & $^{1}$S$_{0}$ &$3.921(25)\times10^{-40}$~\cite{yu_finite-field_2013}
	  \end{tabular}
	\end{ruledtabular}
      \end{table}
      \begin{table}[htpb]
	\caption{\label{tab:examples} Exemplary set of experimental parameters for the proposed experiments using a moving optical lattice with a wavelength of $\unit[1064]{nm}$ and $\unit[200]{mW}$ of power in each beam focussed to a beam waist of $\unit[50]{\upmu m}$}.
	\begin{ruledtabular}	
	\begin{tabular}{@{}lll@{}}
	Parameter  &Example 1 & Example 2\\
      reference &$^{4}$He$^{+}$ $\left[^{2}\mathrm{S}_{1/2}\right]$ & $^{9}$Be$^{+}$ $\left[^{2}\mathrm{S}_{1/2}\right]$ \\
      target &$^{9}$Be$^{+}$ $\left[^{2}\mathrm{S}_{1/2}\right]$ & $^{27}$Al$^{+}$ $\left[^{1}\mathrm{S}_{0}\right]$ \\
      $\eta_\mathrm{IP}^{(r)}$ & $0.14$ & $0.08$ \\
      $\eta_\mathrm{IP}^{(t)}$ & $0.20$ & $0.12$\\
      $\eta_\mathrm{OP}^{(r)}$ & $0.21$ & $0.14$\\
      $\eta_\mathrm{OP}^{(t)}$ &$-0.06$ & $-0.03$\\
      $\omega_{\mathrm{IP}}$ & $\unit[1.567]{MHz}$ & $\unit[1.531]{MHz}$\\
      $\omega_{\mathrm{OP}}$ & $\unit[3.138]{MHz}$ & $\unit[3.392]{MHz}$\\
      $t_{\mathrm{pulse}}$ & $\unit[6.5]{\upmu s}$& $\unit[6.5]{\upmu s}$\\
      $n_{\mathrm{IP}}$ & $10$& $10$\\
      $n_{\mathrm{OP}}$ & $31$& $17$\\
      $n_{\mathrm{IP}}|\aIP|$ & $1.05$& $0.98$\\
      $n_{\mathrm{OP}}|\aOP|$ & $1.03$& $1.00$
	\end{tabular}
      \end{ruledtabular}
      \end{table}

      In the following, we will provide numbers for two different exemplary combinations of atomic species.
\begin{figure}[htpb]
	\centering
	\includegraphics{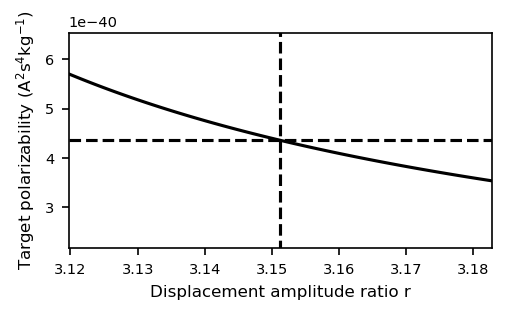}
	\caption{Polarizability determination with $^4$He$^+$ as the reference ion and $^9$Be$^+$ as the target ion. The graph shows the inferred polarizability for a simulated ratio of displacements between the IP and the OP mode. The horizontal dashed line corresponds to the theoretically calculated polarizability of \Be and the vertical dashed line indicates the corresponding displacement amplitude ratio $r$.}
	\label{fig:BeHe}
      \end{figure}

      In the first example $^4$He$^{+}$ is used as the reference ion to measure the polarizability of the $^{9}$Be$^{+}$ $^{2}$S$_{1/2}$ ground state with respect to the helium groundstate. 
      Simultaneous trapping of helium and beryllium ions has been demonstrated already fifteen years ago~\cite{roth_sympathetic_2005} and currently new experiments for the investigation of single helium ions that are sympathetically cooled by beryllium are developed~\cite{mooser_new_2018,krauth_paving_2019}
      The singly charged helium ion is a hydrogen like system and therefore the polarizability of its electronic states can be computed to a high level of precision~\cite{szmytkowski_static_2016}. 
      This advantage comes at the price of a small polarizability compared to the target ion, leading to an uncertainty propagation factor of $23.4$, which means that the determination of the ratio of displacements in terms of fractional accuracy has to be better by this factor to reach the targeted accuracy for the polarizability determination. 
      It should however been mentioned that state of the art polarizability measurements are limited to the few permille or even percent level.
      In figure \ref{fig:BeHe} the dependence of the inferred displacement ratio on the actual value of the beryllium ground state polarizability is shown. 
      As shown in the supplementary material, in equation~\ref{eq:main} the upper sign applies for this particular example.
      A set of possible experimental parameters is given in table~\ref{tab:examples}. 
      Trapping frequencies of $\omega_{\mathrm{IP}}=\unit[1.567]{MHz}$ and $\omega_{\mathrm{OP}}=\unit[3.138]{MHz}$ result in $\Phi=0$ for a moving optical lattice with $\unit[1064]{nm}$ wavelength. 
      Using $\unit[200]{mW}$ power in each beam focussed to a beam waist of $\unit[50]{\upmu m}$, a convenient choice for the number of pulses is $n_{\mathrm{IP}}=10$ and $n_{\mathrm{OP}}=31$ with pulse length $t_{\mathrm{pulse}}=\unit[6.5]{\upmu s}$ give experimental displacements of $|\taIP|=1.05$ and $|\taOP|=1.03$ which can both be efficiently measured with the techniques described in reference~\cite{wolf_motional_2019}.

%

\begin{figure}[htpb]
	\centering
	\includegraphics{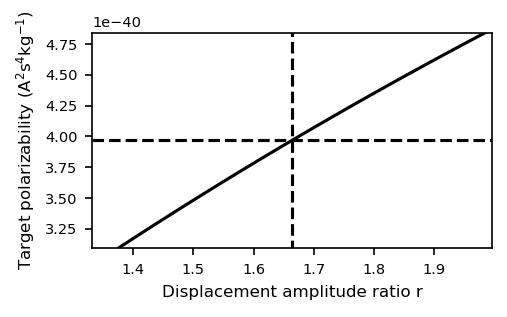}
	\caption{Polarizability determination with \Be as the reference ion and $^{27}$Al$^{+}$ as the target ion. The graph shows the inferred polarizability for a simulated ratio of displacements between the IP and the OP mode. The horizontal dashed line corresponds to the theoretically calculated polarizability of \Al and the vertical dashed line indicates the corresponding displacement amplitude ratio $r$.}
	\label{fig:BeAl}
      \end{figure}

The second example is to use \Be as a reference and to determine the polarizability of the \Al ground state \Snull.
This ion combination has been used in the first demonstration of an aluminum quantum logic clock~\cite{schmidt_spectroscopy_2005,rosenband_frequency_2008}, which is the predecessor of today's most accurate optical clock~\cite{brewer_27+_2019}.
Compared to the previous example the uncertainty propagation is much smaller but with a value of $K(\mu_\mathrm{Al,Be},R_{\mathrm{Al,Be}})=1.2$ still larger than one.
An exemplary set of parameters is given in table~\ref{tab:examples}.
A convenient choice for the trapping frequencies to ensure $\Phi=0$ is $\omega_\mathrm{IP}=\unit[1.531]{MHz}$ and $\omega_\mathrm{OP} = \unit[3.392]{MHz}$.
Using the same optical lattice as previously described, $n_\mathrm{IP}=10$ and $n_\mathrm{OP}=17$ pulses of length $t_\mathrm{pulse} = \unit[6.5]{\upmu s}$ result in displacements on the order of $|\taIP|=0.98$ and $|\taOP|=1.00$, respectively. 
The resulting dependence of the inferred polarizability from the measured ratio $r$ is shown in figure~\ref{fig:BeAl}.

Combining the results of both proposed experiments would allow to build a chain to transfer the accuracy of the polarizability for the helium ground state to the aluminium clock ion.  

Note that the excitation rate for the OP transition in the beryllium-helium system, which is the smallest rate discussed here, is still on the order of \unit[5000]{phonons/s} which is more than three orders of magnitude higher than heating rates in state-of-the-art macroscopic ion traps~\cite{brownnutt_ion-trap_2015}. Moreover, effects of heating should be isotropic in phase space, thereby averaging out in displacement amplitude measurements. Nevertheless, when pursuing experiments aiming for precision below the permille level it is certainly advisable to investigate the process of heating on the actually employed displacement amplitude measurement. As long as the heating rates are constant, potential higher order effects may be mitigated by numerical simulations or calibration measurements. 

%
In summary, we have proposed a novel approach to experimentally determine polarizabilities in a quantum logic scheme. 
	    In contrast to most alternative polarizability measurements, it does not require precise knowledge of the laser intensity, because the co-trapped ion serves as a reference.
	      Subject of future research will be a detailed error analysis to infer experimental limitations of this scheme.
	      Furthermore, it should be mentioned that the systematic uncertainty of an atomic clock due to blackbody radiation depends on the differential polarizability of the involved clock states. Therefore, the described measurement either has to be performed for both clock states or an extensions of the presented scheme has to be developed the differential polarizabilies is directly measured. This could be realized by changing the internal states of the ions during the lattice interrogation, while switching the direction of the displacement, implementing a variant of the quantum lock-in amplifier~\cite{kotler_single-ion_2011}. An alternative route might be adding more target ions prepared in different electronic states. Both extensions are subjects to future research.

  \begin{acknowledgments}
    I thank Piet O. Schmidt, Johannes Kramer and Maximilian J. Zawierucha for valuable discussions.
This work has been funded by the Deutsche Forschungsgemeinschaft (DFG, German Research Foundation) through CRC 1227 (DQ-mat), project B05 with partial support from Germanys Excellence Strategy EXC-2123 QuantumFrontiers 390837967		
  \end{acknowledgments}
%

\appendix
\end{document}


\title{Supplementary Material: A scheme for quantum-logic based transfer of accuracy in polarizability measurements for trapped ions using a moving optical lattice}
\author{Fabian Wolf}
\maketitle
\section{Displacement of motion for ion chains with a moving optical lattice}
A moving optical lattice is formed by two counter-propagating laser beams with a relative frequency detuning $\delta=\omega_2-\omega_1$ and electric field amplitudes $\varepsilon_i$ and direction $\mathbf{u}_i$.
The displacement on mode $k$ due to the interaction of a moving optical lattice with an ion chain is given by
   \begin{equation}
\alpha_k=\frac{\ii}{\hbar} \sum_j \eta_k^{(j)} \varepsilon_1\varepsilon_2^* \alpha^{(j)}(\omega,\mathbf{u}_1,\mathbf{u}_2)\ee^{-\ii\phi_j}t_\mathrm{F}
\label{eq:Displacement}
      \end{equation}
      with the electric field amplitudes $\varepsilon$ and relative phase $\phi_j$ of the laser beams at the position of the $j$-th ion, Planck's constant $\hbar$, the interaction time of the moving optical lattice $t_F$, the dynamic polarizability of the $j$-th ion $\alpha^{(j)}(\omega,\mathbf{u}_1,\mathbf{u}_2)$ and where the Lamb-Dicke parameter
\begin{align}
  \etajk=\bm{k}_{\mathrm{eff}}\cdot \bm{\beta}^{(j)}_k \sqrt{\frac{\hbar}{2m_j\omega_k}}
  \label{eq:LDP}
\end{align}
was introduced. Here, $\bm{k}_{\mathrm{eff}}$ is the effective wave vector. In the case of a single photon transition, this is the wave vector of the exciting field, in case of a Raman transition (which is the case for a moving optical lattice) it is the differential wavevector. In general it is associated with the momentum that is transferred from the light field to the atom.
Usually only the amplitude of the displacement is measured which is given by
\begin{align}
  |\alpha_k|&=\sqrt{\alpha_k\alpha_k^*}\\
  &= \frac{\ii t_\mathrm{F}}{\hbar}\varepsilon_1\varepsilon_2 \sqrt{\left(\sum_j\eta_k^{(j)}\alpha^{(j)}(\omega,\mathbf{u}_1,\mathbf{u}_2)\ee^{-\ii\phi_j}\right)\left(\sum_j\eta_k^{(j)}\alpha^{(j)}(\omega,\mathbf{u}_1,\mathbf{u}_2)\ee^{\ii\phi_j}\right)}\\
  &= \frac{t_\mathrm{F}}{\hbar}\varepsilon_1\varepsilon_2 \sqrt{\sum_{j,l} \eta_k^j \eta_k^l \alpha^i{(j)} (\omega,\mathbf{u}_1,\mathbf{u}_2)\alpha^{(l)}(\omega,\mathbf{u}_1,\mathbf{u}_2)\ee^{-i\left( \phi_j-\phi_l \right)}}
  \label{eq:disp_amp}
\end{align}
The relative phase between the ions $\phi_j-\phi_l$ is given by the ions distance and the projection of the effective wave vector of the laser
\begin{align}
  \phi_j-\phi_l = (\bm{x}_j-\bm{x}_l)\cdot \bm{k}_{\mathrm{eff}}
  \label{<+label+>}
\end{align}
For two and three ion linear chains with identical charge $Z$, the equilibrium positions can be calculated analytically~\cite{morigi_two-species_2001, james_quantum_1998}.
Equation~\ref{eq:disp_amp} can be simplified under the assumption that the relative laser phase is the same at the position of all ions (then, we can set it to $\phi_j=0$ for all $j$ without loss of generality) to
\begin{align}
|\alpha_k| &= \frac{ t_\mathrm{F}}{\hbar}\varepsilon_1\varepsilon_2 \sqrt{\sum_{j,l} \eta_k^{(j)} \eta_k^{(l)} \alpha^{(j)} (\omega,\mathbf{u}_1,\mathbf{u}_2)\alpha^{(l)}(\omega,\mathbf{u}_1,\mathbf{u}_2)}\\
&=\frac{\ii t_\mathrm{F}}{\hbar}\varepsilon_1\varepsilon_2\left|\sum_j\etajk \alphaj\right|
  \label{<+label+>}
\end{align}
which can be achieved either by tuning the trapping frequency and the incident angle and frequency of the laser.
This equation cannot be easily inverted because of the absolute value evaluation on the right side.
Therefore, we introduce the diagonal matrix $\Pi = \delta_{jk}\pi_k=\delta_{jk}\mathrm{sgn}\left(\sum_j \etajk \alphaj \right)$ which results in
\begin{equation}
  \left|\sum_j\etajk \alphaj\right| = \pi_k \sum_j\etajk \alphaj
  \label{<+label+>}
\end{equation}
and gives the relative phase between excitation of different modes i.e. it determines if the force applied by the moving optical lattice pulls the ions towards the high field or low field regions of the light field.
By inverting the matrix equation above, we get an expression, that can be used to infer the polarizabilities from the displacement amplitudes.
\begin{align}
  |\alphaj| = \frac{\hbar}{t_\mathrm{F}\varepsilon_1\varepsilon_2} \left( \etajk \right)^{-1}|\alpha|_k\pi_k
  \label{eq:Matrix_eq}
\end{align}

\subsection{Two-ion chain}
In the following, we will give an explicit formula for the case of a two ion crystal.
The inverse of the Lamb-Dicke matrix is given by
\begin{align}
  \left( \etajk \right)^{-1}&=
  \begin{pmatrix}
    \etarip&\etatip\\
    \etarop&\etatop
  \end{pmatrix}^{-1}\\
  &=
  \frac{1}{\etarip\etatop-\etatip\etarop}
  \begin{pmatrix}
   \etatop & -\etatip\\
   -\etarop& \etarip
  \end{pmatrix}
  \label{<+label+>}
\end{align}
where we labelled the ions $(t)$ for \emph{target} and $(r)$ for \emph{reference} and the modes IP for  \emph{in-phase} and OP for \emph{out-of-phase}.
The Lamb-Dicke parameters and oscillation amplitudes can be derived analytically~\cite{wubbena_sympathetic_2012, morigi_two-species_2001}. Here, we adopt the notation from reference~\cite{clausen_unresolved_2022}.
The mode amplitudes are given by
\begin{align}
  \beta^{(r)}_{\mathrm{IP/OP}}&=\frac{r_{\mathrm{IP/OP}}}{\sqrt{1+r^2_{\mathrm{IP/OP}}}}\\
  \beta^{(t)}_{\mathrm{IP/OP}}&=\frac{1}{\sqrt{1+r^2_{\mathrm{IP/OP}}}}
  \label{<+label+>}
\end{align}
with
\begin{align}
r_{\mathrm{IP}}&=\frac{-\mu+1+\sqrt{\mu^2-\mu+1}}{\sqrt{\mu}}\\
r_{\mathrm{OP}}&=\frac{-\mu+1-\sqrt{\mu^2-\mu+1}}{\sqrt{\mu}}
  \label{<+label+>}
\end{align}
where $\mu=m_{\mathrm{t}}/m_{\mathrm{r}}$ is the mass ratio between target and reference ion.
Combining this with equation~\ref{eq:Matrix_eq} and assuming that the exciting light field intensity and interaction time is kept constant for both excitations we find
\begin{align}
  \frac{|\alphat|}{|\alphar|}=\frac{-\etarop \frac{|\aIP|}{|\aOP|}\pm\etarip}{\etatop \frac{|\aIP|}{|\aOP|}\mp\etatip}
  \label{eq:Moebius_eq}
\end{align}
Defining $r=\frac{|\aIP|}{|\aOP|}$ and inserting the definition of the Lamb-Dicke parameter the previous equation can be written
\begin{align}
  \frac{|\alphat|}{|\alphar|}=\sqrt{\frac{m_t}{m_r}}\frac{-\betarop\sqrt{\frac{\wIP}{\wOP}}r\pm\betarip}{\betatop \sqrt{\frac{\wIP}{\wOP}}r\mp\betatip}
  \label{eq:final}
\end{align}

\subsubsection{How to determine the sign?}
For the case of two ions the matrix $\Pi=\delta_{jk}\pi_k$ that determines the sign in equation~\ref{eq:final} is given by
\begin{align}
  \pi_\mathrm{IP} &= \mathrm{sgn}\left( \etarip \alphar + \etatip \alphat\right)\\
  \pi_\mathrm{OP} &= \mathrm{sgn}\left( \etarop \alphar + \etatop \alphat\right)
  \label{<+label+>}
\end{align}
the upper sign in equation~\ref{eq:final} applies for the case $\pi_\mathrm{IP}=\pi_{\mathrm{OP}}$ and the lower sign for $\pi_\mathrm{IP}=-\pi_{\mathrm{OP}}$.
Without loss of generality we can assume the effective wave vector $\mathbf{k}_{\mathrm{eff}}$ to be positive. In consequence we can choose all Lamb-Dicke parameters to be positive, except for $\etarop$ which gives
\begin{align}
  \pi_\mathrm{IP} &= \mathrm{sgn}\left( |\etarip| \alphar + |\etatip| \alphat\right)\\
  \pi_\mathrm{OP} &= \mathrm{sgn}\left( -|\etarop| \alphar + |\etatop| \alphat\right)
  \label{<+label+>}
\end{align}
We can further write
\begin{align}
  \pi_\mathrm{IP} &= \mathrm{sgn}\left(\alphat + \frac{|\etarip|}{|\etatip|} \alphar\right)\\
  &=\mathrm{sgn}\left( \alphat + \sqrt{\mu}\frac{|\betarip|}{|\betatip|}\alphar \right)\\
  &= \mathrm{sgn}\left( \alphat +\sqrt{\mu}|r_{\mathrm{IP}}|\alphar \right)\\
  &= \mathrm{sgn}\left( \alphat + \left| -\mu+1+\sqrt{\mu^2-\mu+1} \right|\alphar \right)\\
  \pi_\mathrm{OP} &= \mathrm{sgn}\left( \alphat - \frac{|\etarop|}{|\etatop|} \alphar\right)\\
  &= \mathrm{sgn}\left( \alphat - \left| -\mu+1-\sqrt{\mu^2-\mu+1} \right|\alphar \right)
  \label{<+label+>}
\end{align}

\subsubsection{Sensitivity}
Defining $R=|\alphat|/|\alphar|$ we can write
\begin{align}
  R=\Gamma_\mu(r)
  \label{<+label+>}
\end{align}
with
\begin{align}
  \Gamma_\mu(r) = \sqrt{\mu}\frac{\pm\beta_{\mathrm{IP}}^{(r)}-\beta_{\mathrm{OP}}^{(r)}\sqrt{\frac{\omega_{\mathrm{IP}}}{\omega_{\mathrm{OP}}}}r}{\beta_{\mathrm{OP}}^{(t)}\sqrt{\frac{\omega_{\mathrm{IP}}}{\omega_{\mathrm{OP}}}}r\mp\beta^{(t)}_{\mathrm{IP}}}
  \label{<+label+>}
\end{align}
The uncertainty $\Delta r$ in the determination of $r$ can be transferred to final uncertainty in the determination of the target ion polarizability by
\begin{align}
  \Delta |\alphat| = |\alphar|\left.\frac{\partial \Gamma_\mu (r)}{\partial r}\right|_{r=r_0}\Delta r
  \label{dd}
\end{align}
where $r_0$ is the measured ratio of displacement amplitudes.
We can further derive

\begin{align}
  \left.\frac{\partial \Gamma_\mu (r)}{\partial r}\right|_{r=r_0}&:=\Gamma_\mu'(r_0)\\
    &= \Gamma'_\mu\left(\Gamma^{-1}_\mu(R)\right)\\
    &= \frac{1}{\left[\Gamma^{-1}_\mu\right]'(R)}
  \label{<+label+>}
\end{align}
The inverse of $\Gamma_\mu(r)$ and its derivative are given by
\begin{align}
  \Gamma_\mu^{-1}(R)&=\pm\frac{\etatip R+\etarip}{\etatop R + \etarop}\\
  \left[\Gamma_\mu^{-1}\right]'(R)&= \pm \frac{\etarop\etatip-\etarip\etatop}{\left( \etarop+R\etatop \right)^2}\\
  &=\pm \sqrt{\frac{\omega_\mathrm{OP}}{\omega_\mathrm{IP}}}\sqrt{\mu}\frac{\betarop\betatip-\betarip\betatop}{\left( \betarop\sqrt{\mu}+R\betatop\right)^2}
  \label{<+label+>}
\end{align}

The relative error propagates as follows
\begin{align}
  \frac{\Delta|\alphat|}{\alphat}= \underbrace{\frac{\Gamma^{-1}_\mu(R)}{R\left[\Gamma_\mu^{-1} \right]'(R) }}_{K(\mu,R)}\frac{\Delta r}{r}
  \label{<+label+>}
\end{align}
where we defined the sensitivity factor $K(\mu,R)$ which can be written
\begin{align}
  K(\mu,R) &= \frac{\left(\etarip+R\etatip\right)\left( \etarop+R\etatop \right)}{R\left( \etarop\etatip-\etarip\etatop \right)}\\
  &= \frac{\left( \sqrt{\mu}\betarip + R\betatip  \right)\left( \sqrt{\mu}\betarop+R\betatop \right)}{\sqrt{\mu} R \left( \betarop\betatip-\betarip\betatop \right)}\\
  &=\frac{\mu\left(1+2R\right) - R \left( 2+R \right)}{2R \sqrt{\mu^2-\mu+1}}
  \label{<+label+>}
\end{align}

It can be seen that the sensitivity goes to zero if
\begin{align}
  R&=\mu-1+\sqrt{\mu^2-\mu+1}\\
  &= - r_\mathrm{OP}\sqrt{\mu}
  \label{}
\end{align}
which is the condition under which the out-of-phase motion cannot be excited with the chosen phase of the oscillation force set to be equal for each ion.